\documentclass[letterpaper,nofootinbib,prd,amsmath,twocolumn]{revtex4-1}

\usepackage[usenames]{color}
\usepackage{epsfig}
\usepackage{hyperref}

\newcommand\be{\begin{equation}}
\newcommand\ee{\end{equation}}
\newcommand\bea{\begin{eqnarray}}
\newcommand\eea{\end{eqnarray}}
\newcommand\sss{\scriptscriptstyle}
\newcommand\sssfour{\scriptscriptstyle{(4)}}

\begin{document}

\title{Generalized Harmonic Equations in 3+1 Form}
\author{J.~David Brown}
\affiliation{Department of Physics, North Carolina State University,
Raleigh, NC 27695 USA}

\begin{abstract}
The generalized harmonic equations of general relativity are written in 3+1 form. The result is a 
system of partial differential equations with first order time and second order space derivatives 
for the spatial metric, extrinsic curvature, lapse function and shift vector, plus fields that represent 
the time derivatives of the lapse and shift. This allows for a direct 
comparison between the generalized harmonic and the Arnowitt--Deser--Misner formulations. The 3+1 generalized 
harmonic equations are also written in terms of conformal variables and compared to the 
Baumgarte--Shapiro--Shibata--Nakamura equations with moving puncture gauge conditions. 
\end{abstract}
\maketitle

\section{Introduction}
The generalized harmonic equations \cite{Friedrich:GH,Garfinkle:2001ni,Pretorius:2006tp} 
are a symmetric hyperbolic formulation of general relativity. 
They were originally written as a second order system of partial differential equations
for the spacetime metric ${}^{\sssfour}g_{\mu\nu}$. By adding extra variables to represent derivatives 
of ${}^{\sssfour}g_{\mu\nu}$, the generalized harmonic equations can be written 
as a fully first order system \cite{Alvi:2002hu,Lindblom:2005qh},  or as a system with first order time and 
second order space derivatives \cite{Szilagyi:2006qy}. Typically the fundamental variables are the 
components of the spacetime metric and its derivatives. 

In this paper we carry out a 3+1 splitting of the generalized harmonic (GH) equations. In this way the GH system is written 
in terms of traditional 3+1 variables with first--order time and second--order space derivatives. The 
3+1 variables include the spatial metric $g_{ij}$, extrinsic curvature $K_{ij}$, 
lapse function $\alpha$ and shift vector $\beta^i$. The extrinsic curvature is directly related to the time derivative of the 
spatial metric; likewise, we introduce fields $\pi$ and $\rho^i$ that are directly related to the 
time derivatives of $\alpha$ and $\beta^i$. 
The result of this analysis is a concise and elegant expression of Einstein's theory. 

Currently there are two formulations of the Einstein equations in widespread use in the numerical relativity community. 
One is the generalized harmonic system, the other is the Baumgarte--Shapiro--Shibata--Nakamura (BSSN) system 
\cite{Shibata:1995we,Baumgarte:1998te} along with  
moving puncture gauge conditions \cite{Bona:1994dr,Alcubierre:2002kk}. 
The BSSN equations are direct descendants of the Arnowitt--Deser--Misner (ADM) equations, 
which are obtained from a 3+1 splitting of the 
Einstein equations \cite{ADM:Witten}. (See also Refs.~\cite{Alcubierre,BaumgarteShapiro}.) 
ADM and BSSN are typically written as systems with first--order time and second--order space 
derivatives. The fundamental variables for ADM are the 3+1 variables $g_{ij}$, $K_{ij}$, 
$\alpha$ and $\beta^i$. BSSN is obtained from a change of variables, defined by conformal splitting, and the 
introduction of new independent variables, namely, the conformal connection functions. Most often the BSSN system is supplemented 
with the moving puncture gauge conditions which take the form of evolution equations for the lapse function and shift
vector. 

In earlier work, Friedrich and Rendall \cite{Friedrich:2000qv} (see also Ref.~\cite{Moesta})
wrote the generalized harmonic equations in terms of 
3+1 variables $g_{ij}$, $\alpha$ and $\beta^i$. Their motivation was not to compare GH to ADM or BSSN. 
Consequently, the relationship between the GH and ADM or BSSN systems has remained obscure.
In Sec.~III the precise relationship between the 
GH equations and the ADM equations is presented. The relationship between 
the GH equations and the BSSN equations is displayed explicitly in Sec.~V. Note that 
in Refs.~\cite{Bernuzzi:2009ex,Alic:2011gg} the Z4 formulation \cite{Bona:2003fj} of 
general relativity is written in 3+1 form with a conformal splitting, and used to compare Z4 to BSSN.  

In Sec.~2 we review the generalized harmonic formulation of general relativity and discuss its interpretation 
as an initial value problem. In Sec.~3 we write the GH equations in 3+1 form, compare the results to ADM, 
and show that the system is symmetric hyperbolic. Technical details are 
contained in Appendix A. In Sec.~4 the 3+1 GH equations are written in terms of conformal variables. The GH equations 
are compared to BSSN and the moving puncture gauge in Sec.~5. In Appendix B we show that the GH system 
with moving puncture gauge conditions has the same level of hyperbolicity as BSSN with the moving puncture 
gauge. A brief summary is provided in Sec.~6. 
\section{Generalized Harmonic Equations}
Let ${}^{\sssfour}g_{\mu\nu}$ denote the physical spacetime metric with
Christoffel symbols ${}^{\sssfour}\Gamma^\mu{}_{\sigma\rho}$,  covariant derivative $\nabla_\mu$,  and 
Ricci curvature ${}^{\sssfour}R_{\mu\nu}$. Let ${}^{\sssfour}\bar\Gamma^\mu{}_{\sigma\rho}$ denote a background connection; this 
connection might be built from a background metric ${}^{\sssfour}\bar g_{\mu\nu}$. As discussed 
in Ref.~\cite{Brown:2010rya}, the background connection is needed for general covariance. For practical applications 
it would be natural to choose $\bar\Gamma^\mu{}_{\sigma\rho}$ to be the flat connection. If the coordinates 
are interpreted as Cartesian, then the components $\bar\Gamma^\mu{}_{\sigma\rho}$ are zero. 

Now introduce a spacetime vector 
field $H^\mu$, the ``gauge source vector", and define  
\be
	{\cal C}^\mu \equiv H^\mu + ({}^{\sssfour}\Gamma^\mu{}_{\sigma\rho} 
	    - {}^{\sssfour}\bar\Gamma^\mu{}_{\sigma\rho} )g^{\sigma\rho} \ .
\ee
Note that the physical and background connections only appear as the difference ${}^{\sssfour}\Gamma^\mu{}_{\sigma\rho} 
- {}^{\sssfour}\bar\Gamma^\mu{}_{\sigma\rho}$, which transforms as a tensor. 

The generalized harmonic equations are 
\begin{subequations}\label{GHequations}
\bea
	{}^{\sssfour}R_{\mu\nu} - \nabla_{(\mu}{\cal C}_{\nu)} & = & {\color{red} - \kappa \left[ n_{(\mu}{\cal C}_{\nu)} 
	-  {}^{\sssfour}g_{\mu\nu} n^\sigma {\cal C}_\sigma/2 \right] }  \nonumber\\
	& & {\color{blue} + 8\pi G\left[ T_{\mu\nu} - {}^{\sssfour}g_{\mu\nu} T^\sigma_\sigma/2\right] }  \ , \\
	{\cal C}^\mu & = & 0 \ .
\eea
\end{subequations}
The term proportional to Newton's constant $G$ represents the matter content, where $T_{\mu\nu}$ is the matter 
stress--energy--momentum tensor. The matter equations of motion imply the conservation laws $\nabla_\mu T^{\mu\nu}$ = 0. 
The term proportional to the constant $\kappa$ enforces constraint damping; it 
depends on a timelike, future--pointing unit vector field $n^\mu$. Below we assume that this vector field is the unit 
normal to a set of spacelike hypersurfaces $t={\rm const}$. 

The GH equations (\ref{GHequations}) are equivalent to the Einstein equations 
${}^{\sssfour}R_{\mu\nu} = 8\pi G\left[ T_{\mu\nu} - {}^{\sssfour}g_{\mu\nu} T^\sigma_\sigma/2\right]$. This follows trivially by inserting 
Eq.~(\ref{GHequations}b) into Eq.~(\ref{GHequations}a). What makes the GH equations interesting, and useful, is 
their interpretation as an initial value problem. 

Let us define ${\cal M}_\mu \equiv -({}^{\sssfour}G_{\mu\nu} - 8\pi T_{\mu\nu}) n^\nu$ where ${}^{\sssfour}G_{\mu\nu}$
is the Einstein tensor and $n^\mu$ is the unit normal to the $t={\rm const}$ slices. Note that 
${\cal H} \equiv -2{\cal M}_\mu n^\mu$ and ${\cal M}_i$ are the Hamiltonian and momentum constraints, respectively. 

The initial value interpretation of the GH equations relies on two key results. 
The first is obtained by contracting Eq.~(\ref{GHequations}a) with the unit normal $n_\mu$. 
This yields an equation of the form \cite{Lindblom:2005qh,Brown:2010rya}
\begin{equation}\label{keyeqn1}
    \partial_t {\cal C}^\mu = \bigl\{ {\hbox{terms $\sim {\cal M}$, ${\cal C}$, $\partial_i {\cal C}$}} \bigr\}
\end{equation}
where the terms on the right--hand side are proportional to the constraints ${\cal M}_\mu$ and ${\cal C}^\mu$, 
and spatial derivatives of ${\cal C}^\mu$. This equation can be rearranged to show that ${\cal M}^\mu$ 
is proportional to ${\cal C}^\mu$, the time derivative of ${\cal C}^\mu$, and spatial derivatives of ${\cal C}^\mu$. 
It follows that for any solution of the GH equations (\ref{GHequations}), the Hamiltonian and momentum 
constraints ${\cal M}^\mu = 0$ must hold.

The second key result is obtained from the covariant derivative 
of Eq.~(\ref{GHequations}a). After applying the Ricci identity and using the result (\ref{keyeqn1}), we find 
\cite{Lindblom:2005qh,Brown:2010rya}
\begin{equation}\label{keyeqn2}
    \partial_t {\cal M}^\mu = \bigl\{ {\hbox{terms $\sim {\cal M}$, $\partial_i {\cal M}$, 
	${\cal C}$, $\partial_i {\cal C}$, $\partial_i\partial_j {\cal C}$}}\bigr\} \ .
\end{equation}
Together, Eqs.~(\ref{keyeqn1}) and (\ref{keyeqn2}) show that as long as the constraints ${\cal C}^\mu = 0$ 
and ${\cal M}^\mu = 0$ hold at the initial time, then they will continue to hold for all time. 

From the preceding analysis we see that a solution of Einstein's equations can be found by choosing initial data 
that satisfy both sets of constraints, ${\cal C}^\mu = 0$ 
and ${\cal M}^\mu = 0$, at the initial time, and then evolving this data into the future using the 
GH equation (\ref{GHequations}a). 

The generalized harmonic formulation of general relativity is important because the GH equations are 
symmetric hyperbolic, provided the gauge source vector $H^\mu$ is specified directly as a function of the spacetime 
coordinates $x^\mu$ and metric ${}^{\sssfour}g_{\mu\nu}$. 
In particular, the second derivative terms in Eq.~(\ref{GHequations}a) combine to 
form a wave operator ${}^{\sssfour}g^{\sigma\rho}\partial_\sigma \partial_\rho$ acting on the 
spacetime metric ${}^{\sssfour}g_{\mu\nu}$. If, on the other hand, the $H^\mu$'s are specified directly and 
depend on $\partial_\sigma {}^{\sssfour}g_{\mu\nu}$, 
then Eq.~(\ref{GHequations}a) will include terms that interfere with the nice wave operator. In general 
the system will no longer be symmetric hyperbolic. 

In much of the early numerical work with the 
GH equations, $H^\mu$ was not specified directly. 
Rather, $H^\mu$ was elevated to the status of a dynamical variable by introducing 
``driver" equations. With a driver equation, $\nabla_\mu \nabla^\mu H^\nu$ or $\partial_t H^\nu$ is set equal to some function of 
${}^{\sssfour}g_{\mu\nu}$ and $H^\mu$ and their derivatives \cite{Pretorius:2006tp,Lindblom:2007xw}. 
In this case the analysis of hyperbolicity is more complicated. 

Recent work by Szilagyi, Lindblom and Scheel 
\cite{Szilagyi:2009qz} has shown the practical benefits of the ``damped wave gauge". For this gauge condition $H^\mu$ is 
specified directly as a function of the spacetime metric.  Throughout this paper I will assume that the 
gauge source vector is specified directly. If it depends only on the coordinates $x^\mu$ and metric ${}^{\sssfour}g_{\mu\nu}$, then the 
system is symmetric hyperbolic. In Sec.~5 we consider the GH equations with the moving puncture gauge.
In this case $H^\mu$ depends on derivatives of the metric, and the system is not symmetric hyperbolic. 
In Appendix B we show that this system has the same level of hyperbolicity as BSSN with the moving puncture gauge. 

\section{GH equations in 3+1 form}
Let us begin by reviewing the 3+1 decomposition of the Einstein equations 
\cite{ADM:Witten,Alcubierre,BaumgarteShapiro}. The analysis 
yields evolution equations 
\begin{subequations}\label{ADMeqns}
\bea
	\partial_\perp g_{ij} & = & -2\alpha K_{ij} \ ,\\
	\partial_\perp K_{ij} & = & \alpha\left[ R_{ij} - 2K_{ik}K^k_j + K K_{ij} \right]
	- D_i D_j\alpha 
	\nonumber\\ & &  {\color{blue} - 8\pi G\alpha \left[ s_{ij} - g_{ij} (s - \rho)/2 \right] } 
\eea
\end{subequations}
for the spatial metric $g_{ij}$ and extrinsic curvature $K_{ij}$. Here, 
$D_i$ and $R_{ij}$ denote the covariant derivative and Ricci tensor built from the 
spatial metric. 
The lapse function is $\alpha$ and the shift vector is $\beta^i$. The time derivative operator 
is defined by $\partial_\perp \equiv \partial_t - {\cal L}_\beta$, where ${\cal L}_\beta$ is the 
Lie derivative along the shift. 
The matter variables are the energy density $\rho \equiv n^\mu n^\nu T_{\mu\nu}$,  
momentum density $j_i \equiv -n^\mu T_{\mu i}$, and spatial stress $s_{ij} \equiv T_{ij}$. 
The 3+1 splitting of the matter conservation equations $\nabla_\mu T^{\mu\nu} = 0$ gives \cite{Smarr:York}
\begin{subequations}\label{ADMmattereqns}
\bea
	\partial_\perp \rho & = & \alpha s^{ij}K_{ij} + \alpha \rho K - \alpha D_i j^i - 2 j^i D_i\alpha \ ,\\
	\partial_\perp j_i & = & \alpha K j_i - s_{ij}D^j\alpha - \rho D_i\alpha - \alpha D^j s_{ij}  \ .
\eea
\end{subequations}
The spatial metric and extrinsic curvature must also satisfy the Hamiltonian and momentum constraints,
\begin{subequations}\label{ADMconstraints}
\bea
	{ {\cal H}} & \equiv & { K^2 - K_{ij}K^{ij} + R } {\color{blue} - 16\pi G \rho}  = 0 \ ,\\
	{ {\cal M}_i} & \equiv & {  D_jK^j_i - D_i K } {\color{blue} - 8\pi G j_i }  = 0 \ .
\eea
\end{subequations}
If the constraints hold at the initial time, then the evolution equations (\ref{ADMeqns}) 
and (\ref{ADMmattereqns}) insure that they will continue to hold at future times. 

In the numerical relativity 
community the results (\ref{ADMeqns}) are referred to as the Arnowitt--Deser--Misner (ADM) 
equations \cite{ADM:Witten}. Here we use the common convention of writing these equations in the form used by Smarr and 
York \cite{Smarr:1977uf,Smarr:York}.

The mathematical details of the 3+1 splitting of the generalized harmonic equations (\ref{GHequations}) are 
presented in Appendix A. The result is the following system of evolution equations,
\begin{subequations}\label{3+1GHeqns}
\bea
	\partial_\perp g_{ij} & = & -2\alpha K_{ij}  \ , \\
	\partial_\perp K_{ij} & = & \alpha\left[ R_{ij} - 2K_{ik}K^k_j - \pi K_{ij} \right] - D_i D_j\alpha
	\nonumber\\ & &  - \alpha D_{(i}{\cal C}_{j)} 
	{\color{red} - \kappa \alpha g_{ij} {\cal C}_\perp/2 } 
	\nonumber\\ & &  {\color{blue} - 8\pi G\alpha \left[ s_{ij} - g_{ij} (s - \rho)/2 \right] } \ ,   \\
	\partial_\perp\alpha & = &  \alpha^2 \pi - \alpha^2  H_\perp  \ , \\
	\partial_t\beta^i & = & \beta^j\bar D_j\beta^i + \alpha^2\rho^i - \alpha D^i\alpha
		+ \alpha^2  H^i  \ , \\
	\partial_\perp \pi & = & -\alpha K_{ij}K^{ij} + D_i D^i\alpha 
		+ {\cal C}^i D_i\alpha \nonumber\\ & &  {\color{red} -  \kappa \alpha{\cal C}_\perp/2 } 
	{\color{blue} - 4\pi G\alpha (\rho + s) } \ , \\
	\partial_\perp \rho^i & = &  g^{k\ell} \bar D_k \bar D_\ell \beta^i + \alpha D^i\pi - \pi D^i\alpha - 2K^{ij} D_j\alpha
		 \nonumber\\ & & 
		 + 2\alpha K^{jk}\Delta\Gamma^i{}_{jk}
		{\color{red} + \kappa\alpha {\cal C}^i } {\color{blue} - 16\pi G\alpha j^i } \ ,
\eea
\end{subequations}
and constraints,
\begin{subequations}\label{TheConstraints}
\bea
	{\cal C}_\perp & \equiv & \pi + K  \ , \\
	{\cal C}^i & \equiv & -\rho^i + \Delta\Gamma^i{}_{jk}g^{jk}  \ , \\
	{ {\cal H}} & \equiv & { K^2 - K_{ij}K^{ij} + R } {\color{blue} - 16\pi G \rho} \ , \\
	{ {\cal M}_i} & \equiv & {  D_jK^j_i - D_i K } {\color{blue} - 8\pi G j_i } \ .
\eea
\end{subequations}
The dependent variables include the spatial metric $g_{ij}$, extrinsic curvature $K_{ij}$, 
lapse function $\alpha$ and shift vector $\beta^i$. We have also introduced the variables 
$\pi$ and $\rho^i$. Equation (\ref{3+1GHeqns}c) shows that $\pi$ is related to the time 
derivative of $\alpha$. Likewise, from Eq.~(\ref{3+1GHeqns}d) we see that $\rho^i$ is  
related to the time derivative of $\beta^i$. Note that the gauge source vector $H^\mu$ appears in 
these equations as a spatial scalar $H_\perp$ and a spatial vector $H^i$. The source $H_\perp$ 
appears in the evolution equation (\ref{3+1GHeqns}c) for the lapse $\alpha$, while the source $H^i$ 
appears in the evolution equation (\ref{3+1GHeqns}d) for the shift $\beta^i$. 

In deriving the 3+1 GH equations (\ref{3+1GHeqns}), (\ref{TheConstraints}) we have assumed that the 
only non vanishing components of the background connection ${}^{\sssfour}\bar\Gamma^\mu{}_{\sigma\rho}$ 
are the spatial components ${}^{\sssfour}\bar\Gamma^i{}_{jk}$. This is equivalent to building the 
background connection from a background metric ${}^{\sssfour}\bar g_{\mu\nu}$ which, under a 3+1 
splitting, has unit lapse, vanishing shift, and a time--independent spatial metric. 
In this case the only remaining background structure is the spatial connection whose 
components are 
$\bar\Gamma^i_{jk}  \equiv {}^{\sssfour}\bar\Gamma^i{}_{jk}$. We also assume that the background spatial 
connection is flat, and in Eqs.~(\ref{3+1GHeqns}), (\ref{TheConstraints}) use the notation 
\begin{equation}
    \Delta\Gamma^i{}_{jk} \equiv \Gamma^i_{jk} - \bar\Gamma^i{}_{jk}  \ .
\end{equation}
Finally, we let $\bar D_i$ denote the covariant derivative built from the background connection. 

Comparing the 3+1 GH equations (\ref{3+1GHeqns}a) and (\ref{3+1GHeqns}b) with the ADM equations (\ref{ADMeqns}), 
we find 
\begin{subequations}\label{CompareGHandADM}
\bea
    (\partial_\perp g_{ij})_{\sss GH} - (\partial_\perp g_{ij})_{\sss ADM} & = & 0 \ ,\\
    (\partial_\perp K_{ij})_{\sss GH} - (\partial_\perp K_{ij})_{\sss ADM} & = &  -\alpha {\cal C}_\perp K_{ij} 
        - \alpha D_{(i}{\cal C}_{j)} \nonumber\\ 
        & & {\color{red} - \kappa \alpha g_{ij} {\cal C}_\perp/2 } \ .
\eea
\end{subequations}
As expected, the difference is proportional to the constraints (\ref{TheConstraints}). 

The constraint evolution system for the 3+1 generalized harmonic equations is
\begin{subequations}
\bea
	\partial_\perp {  {\cal C}_\perp} & = & -\alpha K {  {\cal C}_\perp} + \alpha {  {\cal H}}
		+ {  {\cal C}^i} D_i\alpha - \alpha D_i{  {\cal C}^i} \nonumber\\ & & 
		{\color{red} - 2\kappa\alpha{\cal C}_\perp}  \ , \\
	\partial_\perp {  {\cal C}_i} & = & { {\cal C}_\perp} D_i\alpha 
		- \alpha D_i{ {\cal C}_\perp} - 2\alpha { {\cal M}_i} 
		- 2\alpha K_{ij}{ {\cal C}^j}  \nonumber\\ & & 
		{\color{red} -\kappa\alpha {\cal C}_i} \ , \\
	\partial_\perp{ {\cal H}} & = & -2\alpha \pi { {\cal H}} + 2\alpha R{ {\cal C}_\perp}
		 - 4 { {\cal M}_i}D^i\alpha \nonumber\\
		& & - 2\alpha D^i{  {\cal M}_i}  + 2\alpha(K^{ij} - K g^{ij})D_i{ {\cal C}_j} 
		\nonumber\\ & & 
		{\color{red} -2\kappa\alpha {\cal C}_\perp}
		{\color{blue} - 32\pi G \alpha \rho \, {\cal C}_\perp } \ , \\
	\partial_\perp{ {\cal M}_i} & = & - { {\cal H}} D_i\alpha 
		+ (K\delta_i^j - K_i^j)D_j(\alpha { {\cal C}_\perp} )
		- \frac{1}{2}\alpha D_i{ {\cal H}} \nonumber\\
		& & - \alpha\pi{  {\cal M}_i} 
		+ D^j\alpha D_{[i} { {\cal C}_{j}}{}_]
		+ D_i ( \alpha D_j{ {\cal C}^j} )  \nonumber\\
		& & - \frac{1}{2}\alpha R_{ij}{ {\cal C}^j} - \alpha D^jD_j { {\cal C}_i}
		\nonumber\\ & & 
		{\color{red} + \kappa D_i(\alpha {\cal C}_\perp)} 
		 {\color{blue} - 8\pi G\alpha j_i \, {\cal C}_\perp } \ .
\eea
\end{subequations}
These results are found from the evolution equations (\ref{3+1GHeqns}) and (\ref{ADMmattereqns}) 
applied to the definitions  (\ref{TheConstraints}).

The GH equations are symmetric hyperbolic. This can be shown by considering the second--order 
system (\ref{GHequations}a), or the fully first order system of Refs.~\cite{Alvi:2002hu,Lindblom:2005qh}. Gundlach and 
Mart\'{i}n--Garc\'{i}a \cite{Gundlach:2005ta} have given a definition of symmetric hyperbolicity that applies to quasilinear 
systems of partial differential equations with first--order time and second--order space derivatives. 
We can apply their definition to the 3+1 GH equations (\ref{3+1GHeqns}). 

To begin, we assume that the matter fields are not derivatively coupled to gravity; that is, the matter 
Lagrangian does not contain derivatives of the metric. Then the matter variables $\rho$, $j^i$, 
and $s_{ij}$ do not contain derivatives of the gravitational variables 
$g_{ij}$, $K_{ij}$, $\alpha$, $\pi$, $\beta^i$, and $\rho^i$. We also assume, as discussed in 
Sec.~2, that the gauge sources $H^i$ and $H_\perp$ are directly specified in terms of the spacetime coordinates 
and the metric variables $g_{ij}$, $\alpha$, and $\beta^i$, not on their derivatives. 

The analysis can be described as follows. 
In effect, we assign weight $0$ to the metric variables and weight $1$ to the ``velocities" $K_{ij}$, $\pi$, and $\rho^i$.
One unit of weight is added for each derivative. 
We introduce the weight $1$ variables
\begin{subequations}
\bea
	g_{mij} & \equiv & \partial_m g_{ij} \ , \\
	\alpha_i & \equiv & \partial_i \alpha \ , \\
	\beta_{ij} & \equiv & (\partial_i\beta^k) g_{kj} \ , 
\eea
\end{subequations}
defined as derivatives of the weight $0$ variables, and compute their equations of motion by differentiating 
Eqs.~(\ref{3+1GHeqns}a), (\ref{3+1GHeqns}c) and (\ref{3+1GHeqns}d). 
Note that $\partial_i\alpha_j$, $\partial_i g_{jk\ell}$, and $\partial_i(\beta_{jk}g^{k\ell})$ 
are symmetric in $i$ and $j$. 
The principal parts of the GH equations (\ref{3+1GHeqns}) are constructed from the highest weight terms in the 
equations of motion for the weight $1$ variables; these are
\begin{subequations}\label{PrincPartsEqns}
\bea
	\check\partial_t g_{mij} & \cong & 2\partial_m \beta_{(ij)} - 2\alpha \partial_m K_{ij} \\
	\check\partial_t K_{ij} & \cong & -\frac{1}{2}\alpha g^{mn}\partial_m g_{nij} 
		+ \alpha \partial_{(i}\rho_{j)} - \partial_i \alpha_j \\
	\check\partial_t \alpha_i & \cong & \alpha^2 \partial_i \pi \\
	\check\partial_t \pi & \cong & g^{ij} \partial_i\alpha_j \\
	\check\partial_t \beta_{ij} & \cong & \alpha^2 \partial_i \rho_j - \alpha \partial_i\alpha_j \\
	\check\partial_t \rho_i & \cong & \alpha \partial_i\pi + g^{jk}\partial_j\beta_{ki} \ ,
\eea
\end{subequations}
where $\cong$ denotes equality apart from lower weight terms. Here 
we have defined the operator $\check\partial_t \equiv \partial_t - \beta^k\partial_k$. 

We now define the quadratic form 
\bea\label{conservedenergy}
	\varepsilon & = & M^{ijk\ell} \biggl[ \frac{1}{4} g^{mn} g_{mij}\, g_{nk\ell}
		\nonumber\\ & & \qquad\quad
		+ (K_{ij} - \beta_{(ij)}/\alpha)(K_{k\ell} - \beta_{(k\ell)}/\alpha)\biggr]    \nonumber \\
		& & + M^{ij} \biggl[ \frac{1}{\alpha^2} g^{k\ell} \beta_{ki}\, \beta_{\ell j}
		\nonumber\\ & & \qquad\quad 
		+ (\rho_i - \alpha_i/\alpha)(\rho_j - \alpha_j/\alpha) \biggr] \nonumber\\
		& & + M\left[ (\pi)^2 + \frac{1}{\alpha^2} g^{ij} \alpha_i \,\alpha_j \right] 
\eea
where the tensors $M^{ijk\ell}$, $M^{ij}$ and $M$ (not related to one another) are positive definite. 
Direct calculation using Eqs.~(\ref{PrincPartsEqns}) shows that the time derivative of $\varepsilon$ 
has a principal part that can be written as the gradient of a vector $\phi^i$; that is, 
$\partial_t\varepsilon \cong \partial_i \phi^i$. This shows that 
$\varepsilon$ is a quadratic, positive--definite energy density with flux $\phi^i$. 
It follows from the theorems of Gundlach and Mart\'{i}n--Garc\'{i}a \cite{Gundlach:2005ta} that 
the system (\ref{3+1GHeqns}) is symmetric hyperbolic. 

\section{GH equations in conformal variables}
In 3+1 form the GH  variables are $g_{ij}$, $K_{ij}$, $\alpha$, $\pi$, $\beta^i$, $\rho^i$. Introduce 
a time--independent spatial density of weight $2$, denoted $\bar\gamma$. As this notation suggests, $\bar\gamma$ can be chosen as the 
determinant of a background metric $\bar\gamma_{ij}$, and this same background metric can be used to define the background 
connection $\bar\Gamma^k{}_{ij}$.  Now consider the conformal variables $\tilde\gamma_{ij}$, 
$\tilde A_{ij}$, $\varphi$, $ K$, $\tilde\Lambda^i$, $\alpha$, $\pi$, and $\beta^i$ defined by
\begin{subequations}\label{convardefs}
\bea
	\tilde\gamma_{ij} & = & (\bar\gamma/g)^{1/3} g_{ij} \ ,\\
	\tilde A_{ij} & = & (\bar\gamma/g)^{1/3} \left[ K_{ij} - \frac{1}{3} g_{ij} K \right] \ ,\\
	\varphi & = & \frac{1}{12} \ln(g/\bar\gamma) \ ,\\
	\tilde\Lambda^i & = & (g/\bar\gamma)^{1/3}\rho^i + \frac{1}{6} (g/\bar\gamma)^{-2/3} 
	D^i (g/\bar\gamma) \ .
\eea
\end{subequations}
Note that the determinant of $\tilde\gamma_{ij}$ is $\bar\gamma$ and the trace of $\tilde A_{ij}$ 
vanishes.  The inverse relations are 
\begin{subequations}
\bea
	g_{ij} & = & e^{4\varphi}\tilde\gamma_{ij} \ ,\\
	K_{ij} & = & e^{4\varphi}\left(\tilde A_{ij} + \frac{1}{3} \tilde\gamma_{ij} K \right) \ ,\\
	\rho^i & = & e^{-4\varphi}\tilde\Lambda^i - 2e^{-4\varphi}\tilde\gamma^{ij} \partial_j\varphi  \ .
\eea
\end{subequations}
Indices on the new variables $\tilde A_{ij}$ and $\tilde\Lambda^i$ are raised and lowered with the conformal 
metric $\tilde\gamma_{ij}$. 

In terms of the conformal variables, the GH equations (\ref{3+1GHeqns}) are 
\begin{widetext}
\begin{subequations}\label{CGHeqnssimple}\allowdisplaybreaks
\bea
	\partial_\perp \tilde\gamma_{ij} & = & -\frac{2}{3} \tilde\gamma_{ij} \bar D_k \beta^k 
		-2\alpha \tilde A_{ij}  \ ,\\
	\partial_\perp \varphi & = & \frac{1}{6} \bar D_k \beta^k - \frac{1}{6}\alpha  K  \ ,\\
	\partial_\perp  K & = &  \alpha\tilde A_{ij} \tilde A^{ij} 
		+ \frac{1}{3} \alpha  K^2 
		- e^{-4\varphi} \left[ \tilde D^2 \alpha + 2 \tilde D^i \varphi \tilde D_i\alpha \right] 
		+ \alpha \left( {\cal H} - K{\cal C}_\perp - \tilde D_i{\cal C}^i 
		- 6{\cal C}^i\partial_i\varphi \right) \nonumber\\
		& & 
		{\color{red} -\,3\alpha \kappa {\cal C}_\perp/2} {\color{blue}\, +\, 4\pi G\alpha (\rho + s)} \ ,\\
	\partial_\perp \tilde A_{ij} & = & e^{-4\varphi}\biggl[ \alpha \tilde{\cal R}_{ij} 
		- 2\alpha\tilde D_i\tilde D_j\varphi + 4\alpha\tilde D_i\varphi \tilde D_j\varphi 
		- \tilde D_i \tilde D_j\alpha + 4\tilde D_{(i} \alpha \tilde D_{j)}\varphi 
		{\color{blue} \, -\, 8\pi G\alpha s_{ij}}  \biggr]^{TF}
		\nonumber\\ & & 
		- \frac{2}{3}\tilde A_{ij} \bar D_k\beta^k 
		-2\alpha \tilde A_{ik} \tilde A^k_j  
		+ \alpha K \tilde A_{ij}  - \alpha {\cal C}_\perp \tilde A_{ij} 
		+ \alpha e^{-4\varphi}\left[ 4{\cal C}_{(i}\tilde D_{j)}\varphi 
		- {\cal C}_k\Delta\tilde\Gamma_{(ij)}{}^k \right]^{TF} \ ,\\
	\partial_\perp\tilde\Lambda^i & = & \tilde\gamma^{k\ell}\bar D_k \bar D_\ell \beta^i 
		 + \frac{2}{3}\tilde\Lambda^i\bar D_k\beta^k 
		+ \frac{1}{3}\tilde D^i(\bar D_k\beta^k) 
		- 2\tilde A^{ik} \partial_k\alpha + 2\alpha \tilde A^{k\ell}\Delta\tilde\Gamma^i{}_{k\ell} 
		+ 12\alpha \tilde A^{ik}\partial_k\varphi
		\nonumber\\ 
		& & - \frac{4}{3} \alpha \tilde D^i K 
		+ \alpha \tilde D^i {\cal C}_\perp + \frac{2}{3} \alpha e^{4\varphi}K {\cal C}^i
		{\color{red}\, +\, \kappa\alpha e^{4\varphi}{\cal C}^i } 
		{\color{blue}\, -\, 16\pi G\alpha e^{4\varphi} j^i} \ ,\\
	\partial_\perp \alpha & = & \alpha^2 \pi - \alpha^2 H_\perp \ ,\\
	\partial_t\beta^i & = & \beta^j \bar D_j\beta^i + \alpha^2 H^i 
		+ \alpha^2 e^{-4\varphi}\left[ \tilde\Lambda^i - 2\tilde D^i\varphi 
		- \tilde D^i\alpha /\alpha \right] \ ,\\
	\partial_\perp \pi & = & -\alpha \tilde A_{ij} \tilde A^{ij} 
		- \frac{1}{3} \alpha  K^2 
		+ e^{-4\varphi} \left( \tilde D^2 \alpha + 2\tilde D^i\varphi \tilde D_i \alpha \right) 
		+ {\cal C}^i\tilde D_i\alpha 
		{\color{red}\, -\, \kappa \alpha {\cal C}_\perp/2} 
		{\color{blue}\, -\, 4\pi G\alpha (\rho + s) } \ ,
\eea
\end{subequations}
where 
\be
	\tilde{\cal R}_{ij} \equiv -\frac{1}{2} \tilde\gamma^{k\ell} \bar D_k \bar D_\ell \tilde\gamma_{ij} 
		+ \tilde\gamma^{k\ell} [\Delta\tilde\Gamma^m{}_{k\ell}\Delta\tilde\Gamma_{(ij)m}
		+ 2\Delta\tilde\Gamma^m{}_{k(i} \Delta\tilde\Gamma_{j)m\ell} 
			+ \Delta\tilde\Gamma^m{}_{ik} \Delta\tilde\Gamma_{mj\ell} ] 
		+ \tilde\gamma_{k(i} \bar D_{j)} \tilde\Lambda^k \ .
\ee
\end{widetext} 
We have also defined
\be
    \Delta\tilde\Gamma^i{}_{jk} = \tilde\Gamma^i{}_{jk} - \bar\Gamma^i_{jk} 
\ee
where $\tilde\Gamma^i_{jk}$ are the Christoffel symbols built from the conformal metric. 

In terms of conformal variables, the constraints (\ref{TheConstraints}) are 
\begin{subequations}\label{ConfConstraints}
\bea
	{\cal C}_\perp & = & \pi +  K \ ,\\
	{\cal C}_i & = & -\tilde\Lambda_i + \Delta\tilde\Gamma_i{}_{jk} \tilde\gamma^{jk} \ ,\\
	{\cal H} & = & \frac{2}{3}  K^2 
		- \tilde A_{ij} \tilde A^{ij} 
		{\color{blue}\, -\, 16\pi G \rho}   \nonumber\\ & & 
		+ e^{-4\varphi} [ \tilde R - 8\tilde D^i\varphi \tilde D_i \varphi - 8 \tilde D^2 \varphi ]  
		 \ ,\\
	{\cal M}_i & = & \tilde D_j \tilde A^j_i - \frac{2}{3} \tilde D_i  K 
		+ 6\tilde A^j_i \tilde D_j\varphi 
		{\color{blue}\, -\, 8\pi G j_i} \ ,
\eea
\end{subequations}
where $\tilde R$ is the Ricci scalar built from the conformal metric $\tilde\gamma_{ij}$. One must remember that 
the indices on ${\cal C}_i$ and ${\cal M}_i$ are raised and lowered with the physical metric $g_{ij}$. Thus, 
for example, ${\cal C}^i = e^{-4\varphi}(-\tilde\Lambda^i + \Delta\tilde\Gamma^i{}_{jk}\tilde\gamma^{jk})$. 

The system (\ref{CGHeqnssimple}) is, of course, symmetric hyperbolic as long as $H_\perp$ and $H_i$ do not depend on 
the weight $1$ variables ($K$, $\tilde A_{ij}$, $\tilde\Lambda^i$, $\pi$) or 
derivatives of the weight $0$ variables ($\tilde\gamma_{ij}$, $\varphi$, $\alpha$, $\beta^i$).
We can confirm this  by defining
\begin{subequations}
\bea
	\tilde\gamma_{mij} & \equiv & \partial_m \tilde\gamma_{ij} \ ,\\
	\tilde\beta_{ij} & \equiv & (\partial_i \beta^k)\tilde\gamma_{kj} \ ,\\
	\alpha_i & \equiv & \partial_i \alpha \ ,\\
	\varphi_i & \equiv & \partial_i \varphi \ ,
\eea
\end{subequations}
and computing the principal parts of the evolutions equations for the weight $1$ variables: 
\begin{widetext} 
\begin{subequations}\allowdisplaybreaks
\bea
	\check\partial_t \tilde\gamma_{mij} & \cong & 2 \partial_m \tilde\beta_{(ij)} 
		- \frac{2}{3}\tilde\gamma_{ij}\tilde\gamma^{k\ell} \partial_m\tilde\beta_{k\ell}
		- 2\alpha \partial_m\tilde A_{ij}  \ ,\\
	\check\partial_t \varphi_m & \cong & \frac{1}{6} \tilde\gamma^{ij}\partial_m \tilde\beta_{ij} 
		- \frac{1}{6} \alpha \partial_m  K  \ ,\\
	\check\partial_t K & \cong & 
		\alpha e^{-4\varphi} \tilde\gamma^{ij} \left[
		\partial_i\tilde\Lambda_j - 8\partial_i\varphi_j - \frac{1}{\alpha}\partial_i \alpha_j\right] \ ,\\
	\check\partial_t \tilde A_{ij} & \cong & 
		- \alpha e^{-4\varphi} \frac{1}{2} \tilde\gamma^{k\ell} \partial_k \tilde\gamma_{\ell ij}
		+ \alpha e^{-4\varphi}\left[ \partial_i \tilde\Lambda_j - 2\partial_i\varphi_j 
		- \frac{1}{\alpha} \partial_i \alpha_j \right]^{TF} \ ,\\
	\check\partial_t \tilde\Lambda_i & \cong & \tilde\gamma^{k\ell}\partial_k \tilde\beta_{\ell i} 
		+ \frac{1}{3} \tilde\gamma^{k\ell} \partial_i \tilde\beta_{k\ell} 
		- \frac{1}{3} \alpha \partial_i\tilde K 
		+ \alpha \partial_i \pi \ ,\\
	\check\partial_t\alpha_i & \cong & \alpha^2 \partial_i\pi \ ,\\
	\check\partial_t\tilde\beta_{ij} & \cong & \alpha^2 e^{-4\varphi} \left[ \partial_i \tilde\Lambda_j 
		-2\partial_i\varphi_j - \frac{1}{\alpha}\partial_i \alpha_j \right] \ ,\\
	\check\partial_t \pi & \cong & e^{-4\varphi} \tilde\gamma^{ij} \partial_i\alpha_j \ .
\eea
\end{subequations}
One can show by explicit calculation that the positive definite energy density
\bea
	\varepsilon & = & M^{ijk\ell} e^{4\varphi} \left[\frac{1}{4} \tilde\gamma^{mn}
		(\tilde\gamma_{mij} + 4\tilde\gamma_{ij}\varphi_m)(\tilde\gamma_{nk\ell} + 4\tilde\gamma_{k\ell}\varphi_n)
		+  e^{4\varphi}(\tilde A_{ij} + \tilde\gamma_{ij} K/3 - \tilde\beta_{(ij)}/\alpha)
		(\tilde A_{k\ell} + \tilde\gamma_{k\ell} K/3 - \tilde\beta_{(k\ell)}/\alpha) \right]
		\nonumber\\ & & 
		+ M^{ij} \left[ \frac{1}{\alpha^2}e^{4\varphi}\tilde\gamma^{k\ell} \tilde\beta_{ki}
			\tilde\beta_{\ell j} 
			+ (\tilde\Lambda_i - 2\varphi_i - \alpha_i/\alpha)
			(\tilde\Lambda_j - 2\varphi_j - \alpha_j/\alpha) \right] 
		\nonumber\\ & & 
		+ M\left[ (\pi)^2 + \frac{1}{\alpha^2}e^{-4\varphi}\tilde\gamma^{ij} \alpha_i\alpha_j \right]
\eea
satisfies the conservation equation $\partial_t\varepsilon \cong \partial_i\phi^i$. 
\end{widetext}

\section{Comparison with BSSN and the moving puncture gauge}
The BSSN equations in covariant form are \cite{Brown:2009dd}
\begin{widetext}
\begin{subequations}\label{BSSNeqns}\allowdisplaybreaks
\bea
	\partial_\perp \tilde\gamma_{ij} & = & -\frac{2}{3} \tilde\gamma_{ij} \bar D_k \beta^k 
		-2\alpha \tilde A_{ij}  \ ,\\
	\partial_\perp \varphi & = & \frac{1}{6} \bar D_k \beta^k - \frac{1}{6}\alpha  K \ , \\
	\partial_\perp  K & = &  \alpha\tilde A_{ij} \tilde A^{ij} + \frac{1}{3} \alpha  K^2 
		- e^{-4\varphi} \left[ \tilde D^2 \alpha + 2 \tilde D^i \varphi \tilde D_i\alpha \right] 
		{\color{blue}\, +\, 4\pi G\alpha(\rho + s) } \ ,\\
	\partial_\perp \tilde A_{ij} & = & e^{-4\varphi}\biggl[ \alpha \tilde{\cal R}_{ij} 
		- 2\alpha\tilde D_i\tilde D_j\varphi + 4\alpha\tilde D_i\varphi \tilde D_j\varphi 
		- \tilde D_i \tilde D_j\alpha + 4\tilde D_{(i} \alpha \tilde D_{j)}\varphi 
		{\color{blue} \, -\, 8\pi G\alpha s_{ij}}  \biggr]^{TF}
		\nonumber\\ & & 
		- \frac{2}{3}\tilde A_{ij} \bar D_k\beta^k 
		-2\alpha \tilde A_{ik} \tilde A^k_j  
		+ \alpha  K \tilde A_{ij}   \ ,\\
	\partial_\perp\tilde\Lambda^i & = & -\tilde\gamma^{jk}{\cal C}_j \bar D_k \beta^i +
		\tilde\gamma^{k\ell}\bar D_k \bar D_\ell \beta^i 
		+ \frac{2}{3}\tilde\gamma^{jk}\Delta\tilde\Gamma^i{}_{jk} \bar D_\ell\beta^\ell
		+ \frac{1}{3}\tilde D^i(\bar D_k\beta^k) 
		- 2\tilde A^{ik} \partial_k\alpha \nonumber \\ & & 
		+ 2\alpha \tilde A^{k\ell}\Delta\tilde\Gamma^i{}_{k\ell} 
		+ 12\alpha \tilde A^{ik}\partial_k\varphi
		- \frac{4}{3} \alpha \tilde D^i  K 
		{\color{blue}\, -\, 16\pi G\alpha e^{4\varphi} j^i}  \ .
\eea
\end{subequations}
\end{widetext}
The variables $\tilde\Lambda^i$ are the ``conformal connection functions". 
If the background is flat and the coordinates are interpreted as Cartesian, then the background connection 
vanishes, $\bar\Gamma^i{}_{jk} = 0$. (We also have $\bar D_i = \partial_i$.) In this case it is common to use the notation 
$\tilde\Gamma^i \equiv \tilde\Gamma^i{}_{jk}\tilde\gamma^{jk}$ for these variables 
rather than $\tilde\Lambda^i$.
Also observe that the first term on the right--hand side of Eq.~(\ref{BSSNeqns}e), 
$-\tilde\gamma^{jk}{\cal C}_j \bar D_k \beta^i 
= \tilde\Lambda^j \bar D_j\beta^i -\tilde\gamma^{k\ell} \Delta\tilde\Gamma^j{}_{k\ell} \bar D_j \beta^i$, 
and the Lie derivative term on the 
left--hand side, $-{\cal L}_\beta \tilde\Lambda^i = -\beta^j \bar D_j \tilde \Lambda^i + \tilde\Lambda^j \bar D_j \beta^i$, 
combine to insure that only derivatives of $\tilde\Lambda^i$, and not $\tilde\Lambda^i$ itself, appear in Eq.~(\ref{BSSNeqns}e). 
This rule is discussed in Ref.~\cite{Alcubierre:2002kk} and is followed by most numerical relativity groups who use the BSSN system. 

The BSSN equations are usually accompanied by the moving puncture gauge conditions,
\begin{subequations}\label{MPGconditions}
\bea
	\partial_t \alpha & = & \beta^i\partial_i\alpha -2\alpha  K \ ,\\
	\partial_t\beta^i & = & \beta^j\bar D_j \beta^i + \frac{3}{4} \tilde\Lambda^i - \eta\beta^i \ ,
\eea
\end{subequations}
where $\eta$ is a parameter, independent of the field variables.  Eqs.~(\ref{MPGconditions}a) and (\ref{MPGconditions}b) are 
the 1+log slicing \cite{Bona:1994dr} and the gamma--driver shift conditions, respectively. 
The gamma--driver shift is often written as a system of first--order equations for 
the shift vector $\beta^i$ and an auxiliary field $B^i$ \cite{Alcubierre:2002kk}. As shown in Ref.~\cite{vanMeter:2006vi}, these equations can be 
integrated to yield the single equation (\ref{MPGconditions}b) for $\beta^i$. 

By explicitly comparing the GH equations in conformal variables,  Eqs.~(\ref{CGHeqnssimple}), 
with the BSSN equations (\ref{BSSNeqns}), we find 
\begin{widetext}
\begin{subequations}\label{Diffeqns}
\bea
	(\partial_t \tilde\gamma_{ij})_{\sss GH} - (\partial_t \tilde\gamma_{ij})_{\sss BSSN} & = & 0 \ ,\\
	(\partial_t \varphi)_{\sss GH} - (\partial_t \varphi)_{\sss BSSN} & = & 0  \ ,\\
	(\partial_t K)_{\sss GH} - (\partial_t K)_{\sss BSSN}  & = & 
		\alpha({\cal H} -K{\cal C}_\perp - \tilde D_i {\cal C}^i - 6{\cal C}^i \partial_i\varphi ) 
		{\color{red}\, -\, 3 \alpha\kappa{\cal C}_\perp/2 } \ ,\\
	(\partial_t \tilde A_{ij})_{\sss GH} - (\partial_t \tilde A_{ij})_{\sss BSSN}  & = & 
	 - \alpha {\cal C}_\perp \tilde A_{ij} 
		+ \alpha e^{-4\varphi}\left[ 4{\cal C}_{(i}\tilde D_{j)}\varphi 
		- {\cal C}_k\Delta\tilde\Gamma_{(ij)}{}^k \right]^{TF}	\ ,\\
	(\partial_t \tilde\Lambda^i)_{\sss GH} - (\partial_t \tilde\Lambda^i)_{\sss BSSN} & = & 
		\tilde\gamma^{jk}{\cal C}_j \bar D_k \beta^i 
		- \frac{2}{3} \tilde\gamma^{ij}{\cal C}_j \bar D_k\beta^k
		+ \alpha \tilde D^i {\cal C}_\perp + 2\alpha K \tilde\gamma^{ij}{\cal C}_j/3
		{\color{red}\, +\, \kappa\alpha \tilde\gamma^{ij}{\cal C}_j }  \ .
\eea
\end{subequations}
\end{widetext}
As expected, the differences between GH and BSSN are proportional to the constraints. 
Note  that the terms proportional to ${\cal C}_\perp$ simply exchange $\pi$ for $-K$; likewise, the 
terms proportional to ${\cal C}_i$ simply exchange $\tilde\Lambda^i$ for $\Delta\tilde\Gamma_{ijk}\tilde\gamma^{jk}$. 
Also observe that only a few of the terms on the right--hand sides of Eqs.~(\ref{Diffeqns}) contribute to the principal 
parts of the equations. In particular, we have
\begin{subequations}
\bea
	(\partial_t K)_{\sss GH} - (\partial_t K)_{\sss BSSN} & \cong & 
		\alpha({\cal H}  - \tilde D_i {\cal C}^i ) \\
	(\partial_t \tilde\Lambda^i)_{\sss GH} - (\partial_t \tilde\Lambda^i)_{\sss BSSN} 
		& \cong &  \alpha \tilde D^i {\cal C}_\perp  \ .
\eea
\end{subequations}
The principal parts of the GH and BSSN equations for $\tilde\gamma_{ij}$, $\varphi$, and $\tilde A_{ij}$ coincide. 

The results (\ref{Diffeqns}) provide a simple prescription for converting a BSSN code into a GH code. First, add the terms on the right--hand 
sides of Eqs.~(\ref{Diffeqns}) to the BSSN equations of motion. Next, add the equation of motion (\ref{CGHeqnssimple}h) for $\pi$. Finally, modify the 
evolution equations for $\alpha$ and $\beta^i$ so that they take the form of Eqs.~(\ref{CGHeqnssimple}f) and (\ref{CGHeqnssimple}g). 

With an appropriate choice of the gauge sources $H_\perp$ and $H_i$, we can adopt moving puncture gauge conditions 
within the generalized harmonic formalism.\footnote{Moving puncture gauge conditions for the Z4 formulation of general relativity have been 
discussed in Refs.~\cite{Bernuzzi:2009ex,Bona:2010is,Alic:2011gg}.}
In terms of conformal variables, we need 
\begin{subequations}\label{HsforMPG}
\bea
    H_\perp & = & \pi + 2K/\alpha \ ,\\
    H^i & = & e^{-4\varphi}\left( -\tilde\Lambda^i + 2\tilde D^i\varphi + \tilde D^i\alpha/\alpha\right) 
        \nonumber\\ & & + \frac{3}{4\alpha^2}\tilde\Lambda^i - \frac{\eta}{\alpha^2}\beta^i \ ,
\eea
\end{subequations}
so that the GH equations (\ref{CGHeqnssimple}f) and (\ref{CGHeqnssimple}g) coincide with the moving puncture equations  
(\ref{MPGconditions}). In terms of the original 3+1 variables, we have
\begin{subequations}\label{HiforMPG}
\bea
    H_\perp & = & \pi + 2K/\alpha \ ,\\
    H^i & = & \frac{3}{4\alpha^2} \left[ (g/\bar\gamma)^{1/3}\rho^i + \frac{1}{6} (g/\bar\gamma)^{-2/3} 
	D^i (g/\bar\gamma) \right]  \nonumber\\ & & 
       -\rho^i + D^i\alpha/\alpha - \frac{\eta}{\alpha^2}\beta^i   \ ,
\eea
\end{subequations}
and the moving puncture gauge conditions read 
\begin{subequations}\label{MPGinGHgravity}
\bea
    \partial_t \alpha & = & \beta^i\partial_i \alpha -2\alpha K \ ,\\
    \partial_t \beta^i & = & \beta^j \bar D_j\beta^i - \eta \beta^i  \nonumber\\ & & 
    + \frac{3}{4} \left[ (g/\bar\gamma)^{1/3}\rho^i + \frac{1}{6} (g/\bar\gamma)^{-2/3} 
	D^i (g/\bar\gamma) \right] \ .\quad
\eea
\end{subequations}
With the moving puncture gauge, the $H$'s depend on weight $1$ variables and derivatives of weight $0$ variables. 
This spoils the symmetric hyperbolicity of the system. 
In Appendix B 
we analyze the GH equations with moving puncture gauge conditions and show that they are strongly hyperbolic 
as long as the condition $2\alpha \ne (g/\bar\gamma)^{1/3}$ is met. 
Note that one can use the constraint ${\cal C}^i = 0$ to exchange 
$\rho^i$ for $\Delta\Gamma^i{}_{jk} g^{jk}$ in Eq.~(\ref{MPGinGHgravity}b). This does not affect the hyperbolicity of the system. 

\section{Summary}
The generalized harmonic equations have been written in 3+1 form using as independent 
variables the spatial metric $g_{ij}$, extrinsic curvature $K_{ij}$, lapse function $\alpha$ and shift vector $\beta^i$, 
as well as fields $\pi$ and $\rho^i$ related to the time derivatives of $\alpha$ and $\beta^i$. 
The resulting set of evolution equations (\ref{3+1GHeqns}) and constraints (\ref{TheConstraints}) are a concise and 
elegant formulation of general relativity. 
The GH evolution system is symmetric hyperbolic with the conserved, positive definite energy density displayed in Eq.~(\ref{conservedenergy}). 

The 3+1 GH equations are written in terms of conformal variables in Eqs.~(\ref{CGHeqnssimple}) and (\ref{ConfConstraints}). This allows for a direct comparison with 
the BSSN formulation of Einstein's theory, and provides a simple prescription for converting a BSSN code into 
a GH code. The moving puncture gauge conditions cannot be used with the GH equations without spoiling symmetric hyperbolicity. 
Nevertheless, the GH system with moving puncture gauge has the same level of hyperbolicity as the BSSN system with moving puncture gauge. 

\begin{acknowledgments}
This work was supported by NSF Grant PHY--0758116 to North Carolina State University.  
I would like to thank Peter Diener, Carlos Lousto and Jeffrey Winicour for helpful questions and comments.  
\end{acknowledgments}
\appendix
\section{3+1 splitting}
In this appendix we derive the equations of motion (\ref{3+1GHeqns}) and the constraints (\ref{TheConstraints}) by carrying 
out a 3+1 splitting of the spacetime generalized harmonic equations (\ref{GHequations}).

Let $n_\mu = -\alpha \delta_\mu^t$ denote the covariant normal to the 
spacelike hypersurfaces $t={\rm const}$; the contravariant form is $n^\mu = (\delta^\mu_t - \beta^i \delta_i^\mu)/\alpha$. 
Also introduce the operator $X^\mu_i = \delta^\mu_i$ that projects spacetime covectors into spatial covectors. The 
covariant form of this operator is $X^i_\mu = \delta^i_\mu + \beta^i \delta_\mu^t$; it satisfies $X_\mu^i X^\mu_j = \delta_j^i$ 
and $X_\mu^i n^\mu = 0$. 

The spacetime metric is written in terms of the normal, spatial projection operator, and spatial metric as 
\be
    {}^{\sssfour}g_{\mu\nu} = g_{ij} X^i_\mu X^j_\nu - n_\mu n_\nu
\ee
Spacetime indices $\mu$, $\nu$, {\em etc.} are 
always raised and lowered with the spacetime metric ${}^{\sssfour}g_{\mu\nu}$ and its inverse ${}^{\sssfour}g^{\mu\nu}$, 
while spatial indices $i$, $j$, {\em etc.} are always raised and lowered 
with the spatial metric $g_{ij}$ and its inverse $g^{ij}$. 

The spacetime Christoffel symbols can be written in terms of 3+1 quantities as 
\begin{subequations}\label{CSsplit}
\bea
    n_\mu  {}^{\sssfour}\Gamma^\mu{}_{\sigma\rho} n^\sigma n^\rho & = & - (\partial_\perp \alpha)/\alpha^2 \ ,\\
    X_\mu^i  {}^{\sssfour}\Gamma^\mu{}_{\sigma\rho} n^\sigma n^\rho & = & (\partial_t \beta^i - \beta^j\partial_j\beta^i)/\alpha^2 
           \nonumber \\ & &  + (D^i \alpha)/\alpha \ ,\\
    n_\mu {}^{\sssfour}\Gamma^\mu{}_{\sigma\rho} X^\sigma_i X^\rho_j & = & K_{ij} \ ,\\
    X_\mu^i {}^{\sssfour}\Gamma^\mu{}_{\sigma\rho} X^\sigma_j X^\rho_k & = & \Gamma^i{}_{jk} \ ,\\
    n_\mu {}^{\sssfour}\Gamma^\mu{}_{\sigma\rho} n^\sigma X^\rho_i & = & -(\partial_i\alpha)/\alpha \ ,\\
    X_\mu^i {}^{\sssfour}\Gamma^\mu{}_{\sigma\rho} n^\sigma X^\rho_j & = & -K^i_j + (\partial_j\beta^i)/\alpha \ ,
\eea
\end{subequations}
where $\partial_\perp \equiv \partial_t - {\cal L}_\beta$ is the time derivative operator used in the main text.
The results (\ref{CSsplit}) are obtained by computing the normal and tangential projections of 
derivatives of the spacetime metric, $\partial_\sigma {}^{\sssfour}g_{\mu\nu}$, and using the definition of the Christoffel 
symbols. Also note that we have used the relation $\partial_\perp g_{ij} \equiv -2\alpha K_{ij}$ that defines 
the extrinsic curvature. This is the equation of motion (\ref{3+1GHeqns}a) for the spatial metric. 

We will also need the splitting of the Ricci tensor, 
\begin{subequations}\label{Riccisplit}
\bea
    {}^{\sssfour}R_{\mu\nu} n^\mu n^\nu & = & (\partial_\perp K + D_i D^i\alpha)/\alpha \nonumber\\ & & 
        - K_{ij} K^{ij} \ ,\\
    {}^{\sssfour}R_{\mu\nu} X^\mu_i X^\nu_j & = & R_{ij} + K K_{ij} - 2K_{ik}K^k_j \nonumber\\ & & 
        -(\partial_\perp K_{ij})/\alpha - (D_i D_j\alpha)/\alpha \ , \quad \\
    {}^{\sssfour}R_{\mu\nu} X^\mu_i n^\nu & = & -D_j K^j_i + D_i K \ ,
\eea
\end{subequations}
and the curvature scalar: 
\bea
    {}^{\sssfour}R & = & R + K_{ij} K^{ij} + K^2 - 2(\partial_\perp K)/\alpha  \nonumber\\ & & 
        - 2(D_i D^i\alpha)/\alpha \ .
\eea
These results can be obtained from the definition of the Riemann tensor in terms of covariant derivatives, 
${}^{\sssfour}R_{\mu\nu\sigma\rho}V^\rho = \nabla_\mu \nabla_\nu V_\sigma - \nabla_\nu \nabla_\mu V_\sigma$, 
or from the definition of Riemann in terms of Christoffel symbols and the results (\ref{CSsplit}). 

The GH constraint ${\cal C}_\mu \equiv H_\mu + ({}^{\sssfour}\Gamma^\mu{}_{\sigma\rho} 
- {}^{\sssfour}\bar\Gamma^\mu{}_{\sigma\rho}) g^{\sigma\rho}$
must be split into a normal component, ${\cal C}_\perp \equiv {\cal C}_\mu n^\mu$, and a tangential component, 
${\cal C}_i \equiv {\cal C}_\mu X^\mu_i$. These calculations depend on the 3+1 splitting of 
the background connection $\bar \Gamma^\mu{}_{\sigma\rho}$. 
Let us assume 
that the background connection is constructed from a background metric ${}^{\sssfour}\bar g_{\mu\nu}$. This 
metric can be split with respect to the $t={\rm const}$ hypersurfaces 
into the 3+1 quantities $\bar g_{ij}$, $\bar\alpha$, and $\bar\beta^i$. 
The results (\ref{CSsplit}), applied to the 
background geometry, can be rearranged to give the components of the background connection:
\begin{subequations}
\bea
    {}^{\sssfour}\bar\Gamma^t{}_{tt} & = & (\partial_t\bar\alpha + \bar\beta^j \partial_j\bar\alpha 
        - \bar\beta^i \bar\beta^j \bar K_{ij})/\bar\alpha \ ,\\
    {}^{\sssfour}\bar\Gamma^t{}_{ti} & = & (\partial_i\bar\alpha - \bar\beta^j \bar K_{ij})/\bar\alpha \ ,\\
    {}^{\sssfour}\bar\Gamma^t{}_{ij} & = & -\bar K_{ij}/\bar\alpha \ ,\\
    {}^{\sssfour}\bar\Gamma^i{}_{tt} & = & \bar\alpha \bar D^i\bar\alpha - 2\bar\alpha \bar\beta^j \bar K_{jk}\bar g^{ki} \nonumber\\ & & 
        - \bar\beta^i(\partial_t\bar\alpha + \bar\beta^j \partial_j\bar\alpha 
        - \bar\beta^j \bar\beta^k \bar K_{jk})/\bar\alpha  \nonumber\\ & & 
        + \partial_t \bar\beta^i + \bar\beta^j \bar D_j \bar \beta^i \ ,\\
    {}^{\sssfour}\bar\Gamma^i{}_{jt} & = & -\bar\beta^i (\partial_j\bar\alpha - \bar\beta^k \bar K_{kj})/\bar\alpha 
        \nonumber\\ & & - \bar\alpha \bar K_{jk}\bar g^{ki} + \bar D_j \bar\beta^i \ ,\\
    {}^{\sssfour}\bar\Gamma^k{}_{ij} & = & \bar\Gamma^k{}_{ij} + \bar\beta^k \bar K_{ij}/\bar\alpha \ .
\eea
\end{subequations}
Here, the background extrinsic curvature is defined by 
$(\partial_t - {\cal L}_{\bar\beta}) \bar g_{ij} \equiv -2\bar\alpha \bar K_{ij}$.

The calculations for the normal and tangential components of the constraint yield
\begin{widetext}
\begin{subequations}\label{rawCperpandCi}
\bea
    {\cal C}_\perp & = & H_\perp + K + \frac{1}{\alpha^2}\partial_\perp\alpha - \frac{\alpha}{\bar\alpha} g^{ij}\bar K_{ij}
        - \frac{1}{\alpha\bar\alpha}   (\partial_t - {\cal L}_{\bar\beta})   \bar\alpha 
        + \frac{1}{\alpha\bar\alpha} \Delta\beta^i \Delta\beta^j \bar K_{ij} 
        + \frac{2}{\alpha\bar\alpha} \Delta\beta^i \partial_i \bar\alpha \ ,\\
    {\cal C}_i & = & H_i + g_{ij}g^{k\ell} \Delta\Gamma^j{}_{k\ell} - \frac{1}{\alpha}\partial_i\alpha 
        - \frac{1}{\alpha^2} g_{ij}(\partial_t\beta^j - \beta^k\bar D_k \beta^j) 
        + \frac{\bar\alpha}{\alpha^2} g_{ij}\bar g^{k\ell} \partial_k\bar\alpha 
        + \frac{1}{\alpha^2} g_{ij}(\partial_t\bar\beta^j - \bar\beta^k \bar D_k\bar\beta^j)  \nonumber\\ & & 
        + \frac{1}{\alpha^2\bar\alpha} g_{ij}\Delta\beta^j \left[ \partial_t\bar\alpha 
            - (2\beta^k - \bar\beta^k)\partial_k\bar\alpha + \alpha^2 g^{k\ell}\bar K_{k\ell}\right] 
        -\frac{1}{\alpha^2\bar\alpha} g_{ij}\Delta\beta^k \left[ \Delta\beta^j \Delta\beta^\ell 
            - 2\bar\alpha^2 \bar g^{j\ell}\right] \bar K_{k\ell} \ .
\eea
\end{subequations}
\end{widetext}
Here we have defined $H_\perp \equiv H^\mu n_\mu$ 
and $\Delta\beta^i \equiv \beta^i - \bar\beta^i$. 

Note that each term in Eq.~(\ref{rawCperpandCi}a) is a spatial scalar, and each term in Eq.~(\ref{rawCperpandCi}b) 
is a spatial covector.  In these equations we can absorb terms that depend
on the physical tensors $g_{ij}$, $\alpha$, $\beta^i$, the background tensors 
$\bar g_{ij}$, $\bar\alpha$, $\bar\beta^i$, and derivatives of these background tensors into 
$H_\perp$ and $H_i$. We cannot absorb terms that depend on derivatives of $g_{ij}$, $\alpha$ or $\beta^i$ because this would 
change the hyperbolicity of the GH system. Thus, we have the following results:
\begin{subequations}
\bea
    {\cal C}_\perp & = & H_\perp + K + \frac{1}{\alpha^2}\partial_\perp \alpha \ ,\\
    {\cal C}_i & = & H_i + \Delta\Gamma_{ijk}g^{jk} - \frac{1}{\alpha}\partial_i\alpha  \nonumber\\ & & 
        -\frac{1}{\alpha^2} g_{ij} (\partial_t \beta^j - \beta^k \bar D_k \beta^j) \ .
\eea
\end{subequations}
Let us define 
\begin{subequations}
\bea
    \pi & \equiv & \frac{1}{\alpha^2}\partial_\perp\alpha + H_\perp \ ,\\
    \rho_i & \equiv & \frac{1}{\alpha^2} g_{ij} (\partial_t \beta^j - \beta^k\bar D_k\beta^j) \nonumber\\ & & 
        + \frac{1}{\alpha}\partial_i\alpha - H_i \ .
\eea
\end{subequations}
When rearranged, these definitions become the equations of motion (\ref{3+1GHeqns}c) and (\ref{3+1GHeqns}d) 
for $\alpha$ and $\beta^i$. The constraints become 
\begin{subequations}\label{constraintsagain}
\bea
    {\cal C}_\perp & = & \pi + K \ ,\\
    {\cal C}_i & = & -\rho_i + \Delta\Gamma_{ijk}g^{jk} \ ,
\eea
\end{subequations}
which are Eqs.~(\ref{TheConstraints}a) and (\ref{TheConstraints}b) from the main text. 

Our next task is to split the terms $\nabla_{(\mu} {\cal C}_{\nu)}$. The normal and tangential projections are 
\begin{subequations}\label{splitDC}
\bea
    n^\mu n^\nu \nabla_{(\mu}{\cal C}_{\nu)} & = & \frac{1}{\alpha} (\partial_\perp {\cal C}_\perp 
        - {\cal C}^i\partial_i\alpha ) \ ,\\
    X^\mu_i X^\nu_j\nabla_{(\mu}{\cal C}_{\nu)} & = & D_{(i}{\cal C}_{j)} + K_{ij} {\cal C}_\perp \ ,\\
    n^\mu X^\nu_i \nabla_{(\mu}{\cal C}_{\nu)} & = & \frac{1}{2\alpha} ( \partial_\perp {\cal C}_i 
        - {\cal C}_\perp \partial_i \alpha) \nonumber\\ & & 
        + \frac{1}{2} \partial_i {\cal C}_\perp + K_{ij} {\cal C}^j \ .
\eea
\end{subequations}
The projections of the spacetime GH equation (\ref{GHequations}a) are obtained from the 
Eqs.~(\ref{Riccisplit}), (\ref{constraintsagain}) and (\ref{splitDC}) above. The result for the 
normal--normal projection is
\bea
    	\partial_\perp \pi & = & -\alpha K_{ij}K^{ij} + D_i D^i\alpha 
		+ {\cal C}^i D_i\alpha \nonumber\\ & &  {\color{red} -  \kappa \alpha{\cal C}_\perp/2 } 
	{\color{blue} - 4\pi G\alpha (\rho + s) }  \ ,
\eea
which is  Eq.~(\ref{3+1GHeqns}e) from the main text. 
The tangential--tangential projection yields
\bea
	\partial_\perp K_{ij} & = & \alpha\left[ R_{ij} - 2K_{ik}K^k_j - \pi K_{ij} \right] - D_i D_j\alpha
	\nonumber\\ & &  - \alpha D_{(i}{\cal C}_{j)} 
	{\color{red} - \kappa \alpha g_{ij} {\cal C}_\perp/2 } 
	\nonumber\\ & &  {\color{blue} - 8\pi G\alpha \left[ s_{ij} - g_{ij} (s - \rho)/2 \right] } \ ,
\eea
which is Eq.~(\ref{3+1GHeqns}b). 

The normal--tangential projection of the spacetime GH equation leads to the result 
\bea
	\partial_\perp \rho^i & = &  g^{k\ell} \bar D_k \bar D_\ell \Delta \beta^i + \alpha D^i\pi 
	    - \pi D^i\alpha - 2K^{ij} D_j\alpha \nonumber\\ & & 
		 + 2\alpha K^{jk}\Delta\Gamma^i{}_{jk}  
		{\color{red} + \kappa\alpha {\cal C}^i } {\color{blue} - 16\pi G\alpha j^i }  \nonumber\\ & & 
		 + \bar g^{ij} g^{k\ell} \bigl[ 2\bar D_k(\bar\alpha \bar K_{j\ell}) 
		    - \bar D_j (\bar\alpha \bar K_{k\ell}) \nonumber\\ & & \qquad\qquad
		    - \Delta\beta^m \bar R_{mk\ell j} \bigr]  \ .
\eea
We now assume the background lapse is unity, $\bar\alpha = 1$, and the background 
shift vanishes, $\bar\beta^i = 0$. We also assume that the background spatial metric $\bar g_{ij}$ 
is flat and time independent. These assumptions imply that the background extrinsic curvature $\bar K_{ij}$ and background Riemann tensor 
$\bar R_{mk\ell j}$ vanish. Then the normal--tangential projection becomes 
\bea
	\partial_\perp \rho^i & = &  g^{k\ell} \bar D_k \bar D_\ell \beta^i + \alpha D^i\pi 
	    - \pi D^i\alpha - 2K^{ij} D_j\alpha \nonumber\\ & & 
		 + 2\alpha K^{jk}\Delta\Gamma^i{}_{jk}  
		{\color{red} + \kappa\alpha {\cal C}^i } {\color{blue} - 16\pi G\alpha j^i }  \ ,
\eea
which is Eq.~(\ref{3+1GHeqns}f) from the main text. 

The analysis shows that the spacetime GH equations (\ref{GHequations}) are equivalent to 
the evolution equations (\ref{3+1GHeqns}) plus the  constraints ${\cal C}_\perp = 0$ and
${\cal C}_i = 0$. 
The constraint evolution system (\ref{TheConstraints}) shows that ${\cal C}_\perp = 0$ and ${\cal C}_i = 0$ will hold for all 
time if and only if all of the constraint functions ${\cal C}_\perp$, ${\cal C}_i$, ${\cal H}$, and ${\cal M}_i$ vanish.  
It is sufficient to impose these constraints at the initial time; the evolution equations will 
insure that they continue to hold into the future. 

\section{Hyperbolicity of the GH equations with moving puncture gauge}
In this section we analyze the hyperbolicity of the generalized harmonic equations with the moving 
puncture gauge conditions (\ref{MPGinGHgravity}). That is, we consider Eqs.~(\ref{3+1GHeqns}) with 
the gauge sources $H_\perp$ and $H_i$ given by Eqs.~(\ref{HiforMPG}). 
Symmetric hyperbolicity is spoiled 
by the presence of $\pi$, $K_{ij}$, $\rho^i$ 
and derivatives of $g_{ij}$ and $\alpha$ in the $H$'s. Nevertheless, the equations form a quasilinear system of partial differential 
equations with first order time and second order space derivatives. We can apply the pseudo--differential 
reduction techniques of Refs.~\cite{Kreiss:2001cu,Nagy:2004td,Kreiss:2006mi}
to check for strong hyperbolicity.

The principal parts of the equations are constructed from
the highest weight terms. We identify the ``coordinate variables"
$g_{ij}$, $\alpha$ and $\beta^i$ as weight 0 and the ``velocity variables" $K_{ij}$, $\pi$ and $\rho^i$ as weight 1.  
Each derivative adds a unit of weight. 
The principal parts  of the GH equations with moving puncture gauge conditions are 
\begin{subequations}\label{PrincParts}
\bea
    \check\partial_t g_{ij} & \cong & 2 g_{k(i}\partial_{j)} \beta^k - 2\alpha K_{ij} \ ,\\
    \check\partial_t K_{ij} & \cong & -\frac{\alpha}{2} g^{k\ell}\partial_k \partial_\ell g_{ij}
        + \alpha \partial_{(i}\rho_{j)} - \partial_i \partial_j \alpha \ ,\\
    \check\partial_t \alpha & \cong & -2\alpha K \ ,\\
    \check\partial_t \beta^i & \cong & \frac{3}{4} (g/\bar\gamma)^{1/3} \left[ \rho^i + \frac{1}{6} 
        g^{ij} g^{k\ell} \partial_j g_{k\ell} \right] \ , \\
    \check\partial_t \pi & \cong & g^{ij} \partial_i \partial_j \alpha \ ,\\
    \check\partial_t \rho^i & \cong & \alpha g^{ij}\partial_j \pi + g^{jk} \partial_j \partial_k \beta^i \ ,
\eea
\end{subequations}
where $\check\partial_t \equiv \partial_t - \beta^i \partial_i $.

Let $n_i$ denote a covector normalized 
by the spatial metric: $n_i g^{ij} n_j = 1$. The principal symbol for the system (\ref{PrincParts}) 
above is defined by 
\begin{subequations}\label{PrincSymb}
\bea
    \check\mu \hat g_{ij}  & = & 2 g_{k(i} n_{j)} \hat\beta^k - 2\alpha \hat K_{ij} \ ,\\
    \check\mu \hat K_{ij} & = & -\frac{\alpha}{2} \hat g_{ij} + \alpha g_{k(i} n_{j)} \hat\rho^k - n_i n_j \hat \alpha \ ,\\
    \check\mu \hat\alpha & = & -2\alpha g^{ij}\hat K_{ij} \ ,\\
    \check\mu \hat\beta^i & = & \frac{3}{4} (g/\bar\gamma)^{1/3} \left[ \hat\rho^i + \frac{1}{6} 
        n^i g^{k\ell} \hat g_{k\ell} \right] \ , \\
    \check\mu \hat \pi & = & \hat\alpha \ ,\\
    \check\mu \hat \rho^i & = & \alpha n^i \hat\pi + \hat \beta^i \ ,
\eea
\end{subequations}
where $\check\mu \equiv \mu - \beta^i n_i$. The proper speed (proper distance per unit proper time as 
measured by observers at rest in the $t={\rm const}$ slices) 
of a characteristic mode is given by $(\beta^\perp-\mu)/\alpha$. (See, for example, the discussion in Ref.~\cite{Brown:2009ki}). 

Now introduce an orthonormal triad consisting of $n_i$ and unit vectors $e^i_A$, with $A = 1, 2$. These vectors 
satisfy $n_i e^i_A = 0$ and $e^i_A g_{ij} e^j_B = \delta_{AB}$. When we project equations (\ref{PrincSymb}) into 
the triad directions $n^i$ and $e^i_A$, the principal symbol separates into blocks that have common transformation properties 
under rotations about the plane orthogonal to $n^i$. The scalar block is 
\begin{subequations}\label{scalarblock}
\bea
    \mu \hat g_{\perp\perp} & = & 2\hat\beta^\perp - 2\alpha \hat K_{\perp\perp} \ ,\\
    \mu \hat g_{AB}\delta^{AB} & = & -2\alpha \hat K_{AB}\delta^{AB} \ ,\\
    \mu \hat K_{\perp\perp} & = & -\frac{\alpha}{2} \hat g_{\perp\perp} + \alpha \hat\rho^\perp - \hat\alpha  \ ,\\
    \mu \hat K_{AB}\delta^{AB} & = & -\frac{\alpha}{2} \hat g_{AB}\delta^{AB} \ ,\\
    \mu \hat\alpha & = & -2\alpha ( \hat K_{\perp\perp} + \hat K_{AB}\delta^{AB} ) \ ,\\
    \mu \hat\beta^\perp & = & \frac{3}{4} (g/\bar\gamma)^{1/3} \hat\rho^\perp  \nonumber\\ & & 
        + \frac{1}{8}(g/\bar\gamma)^{1/3} (\hat g_{\perp\perp} + \hat g_{AB}\delta^{AB} ) \ ,\\
    \mu \hat \pi & = & \hat \alpha \ ,\\
    \mu \hat \rho^\perp & = & \alpha \hat\pi + \hat\beta^\perp \ .
\eea
\end{subequations}
Here and below, the $\perp$ and upper case Latin indices are defined, for example, by 
$\hat g_{\perp\perp} \equiv\hat g_{ij}n^i n^j$ and $\hat g_{AB} \equiv \hat g_{ij} e^i_A e^j_B$. 
The vector block is 
\begin{subequations}\label{vectorblock}
\bea
    \mu \hat g_{\perp A} & = & \hat\beta_A - 2\alpha \hat K_{\perp A} \ ,\\
    \mu \hat K_{\perp A} & = & -\frac{\alpha}{2} \hat g_{\perp A} + \frac{\alpha}{2} \hat\rho_A \ ,\\
    \mu \hat\beta_A & = & \frac{3}{4} (g/\bar\gamma)^{1/3} \hat\rho_A \ ,\\
    \mu \hat\rho_A & = & \hat\beta_A \ .
\eea
\end{subequations}
The tensor block is 
\begin{subequations}\label{tensorblock}
\bea
    \mu \hat g_{AB}^{tf} & = & -2\alpha \hat K_{AB}^{tf} \ ,\\
    \mu \hat K_{AB}^{tf} & = & - \frac{\alpha}{2} \hat g_{AB}^{tf} \ ,
\eea
\end{subequations}
where the superscript $tf$ indicates that the tensor is trace--free in the two--dimensional surface orthogonal to $n_i$. 

A quasilinear system is strongly hyperbolic if its principal symbol possesses a complete set of eigenvectors with
real eigenvalues $\mu$. The tensor block (\ref{tensorblock}) meets these criteria with eigenvalues $\mu = \beta^\perp \pm \alpha$. 
These eigenvalues correspond to proper speeds of $\pm 1$. The vector block also meets the criteria for strong hyperbolicity 
with $\mu = \beta^\perp \pm \alpha$ and $\mu = \beta^\perp \pm \sqrt{3(g/\bar\gamma)^{1/3}}/2$. 
The proper speeds for the vector modes are $\pm 1$ and $\pm \sqrt{3(g/\bar\gamma)^{1/3}}/(2\alpha)$. 

The eigenvalues for the scalar block are 
$\mu = \beta^\perp \pm\alpha$ (with multiplicity two), $\mu = \beta^\perp \pm \sqrt{2\alpha}$, and $\mu = \beta^\perp \pm (g/\bar\gamma)^{1/6}$.
These correspond to proper speeds $\pm 1$ (with multiplicity two), $\pm \sqrt{2/\alpha}$, and $\pm (g/\bar\gamma)^{1/6}/\alpha$. 
The eigenvectors are complete unless the eigenvalues $\beta^\perp \pm \sqrt{2\alpha}$ and
$\beta^\perp \pm (g/\bar\gamma)^{1/6}$ coincide. That is, the scalar block meets the criteria for strong hyperbolicity 
as long as $2\alpha \ne (g/\bar\gamma)^{1/3}$.  

The GH system with moving puncture gauge conditions is strongly hyperbolic everywhere, except for regions of spacetime in which 
$2\alpha = (g/\bar\gamma)^{1/3}$.  This restriction on strong hyperbolicity also applies to BSSN with the 
moving puncture gauge \cite{Beyer:2004sv}. 
In fact, the characteristic speeds for GH with moving puncture gauge are precisely the same as for BSSN with moving puncture gauge. 
It is recognized from studies with the BSSN equations that the condition $2\alpha \ne (g/\bar\gamma)^{1/3}$ is typically violated in 
black hole spacetimes on a 2--dimensional surface in space \cite{Beyer:2004sv,Brown:2009ki}. The breakdown of strong 
hyperbolicity does not appear to cause problems
for finite difference codes. On the other hand, the lack of hyperbolicity can create serious problems for spectral codes that 
rely on the passing of characteristic information between spatial domains \cite{fobssn}.

Recall that the moving puncture gauge conditions (\ref{MPGinGHgravity}) can be modified by using the constraint ${\cal C}^i = 0$ 
to replace $\rho^i$ with $\Delta\Gamma^i{}_{jk} g^{jk}$. With this replacement Eq.~(\ref{PrincParts}d) becomes
\be
    \check\partial_t \beta^i \cong \frac{3}{4}(g/\bar\gamma)^{1/3}\left[ g^{ij}g^{k\ell} \partial_k g_{\ell j} 
        - \frac{1}{3} g^{ij} g^{k\ell} \partial_j g_{k\ell} \right] \ .
\ee
The principal symbol (\ref{PrincSymb}) along with its scalar and vector 
blocks are modified accordingly. However, the eigenvalues are not changed, 
and once again the eigenvectors are complete if $2\alpha \ne (g/\bar\gamma)^{1/3}$.

\bibliography{references}
\end{document}